\documentstyle[aps,prl,graphics,epsfig]{revtex}
\begin{document}

\draft

\twocolumn[\hsize\textwidth\columnwidth\hsize\csname
@twocolumnfalse\endcsname

\title{Conceptual Inadequacy of the Shannon Information in Quantum Measurements}
\author{\v Caslav Brukner and Anton Zeilinger}
\address{Institute for Experimentalphysics, University of Vienna,\\
         Boltzmanngasse 5, A--1090 Vienna, Austria}

\date{\today}

\maketitle

\vskip1pc]

\begin{abstract}

In a classical measurement the Shannon information is a natural
measure of our ignorance about properties of a system. There,
observation removes that ignorance in revealing properties of the
system which can be considered to preexist prior to and
independent of observation. Because of the completely different
root of a quantum measurement as compared to a classical
measurement conceptual difficulties arise when we try to define
the information gain in a quantum measurement using the notion of
Shannon information. The reason is that, in contrast to classical
measurement, quantum measurement, with very few exceptions, cannot
be claimed to reveal a property of the individual quantum system
existing before the measurement is performed.
\end{abstract}


\section{Introduction}

In classical physics information is represented as a binary
sequence, i.e a sequence of bit values, each of which can be
either $1$ or $0$. When we read out information that is carried by
a classical system we reveal a certain bit value that exists even
before the reading of information is performed. For example, when
we read out a bit value encoded as a pit on a compact disk, we
reveal a property of the disk existing before the reading process.

This means that in a classical measurement the particular sequence
of bit values obtained can be considered to be physically defined
by the properties of the classical system
measured\footnote{\label{crv} Even in these cases where classical
physics instead of definite measurement results predicts these
results with certain probabilities, it is still possible at least
in principle, to consider an ensemble of statistically distributed
measurement results as revealing corresponding statistically
distributed properties of the ensemble of classical systems.}. The
information read is then measured by the Shannon measure of
information \cite{shannon} which can operationally be defined as
the number of binary questions (questions with ''yes'' or ''no''
answers only) needed to determine the actual sequence of 0's and
1's.

In quantum physics information is represented by a sequence of
qubits, each of which is defined in a two-dimensional Hilbert
space. If we read out the information carried by the qubit, we
have to project the state of the qubit onto the measurement basis
$\{|0\rangle,|1\rangle\}$ which will give us a bit value of either
0 or 1. Only in the exceptional case of the qubit in an eigenstate
of the measurement apparatus the bit value observed reveals a
property already carried by the qubit. Yet in general the value
obtained by the measurement has an element of irreducible
randomness and therefore cannot be assumed to reveal the bit value
or even a hidden property of the system existing before the
measurement is performed.

This implies that in a sequence of measurements on qubits in a
superposition state $a|0\rangle + b|1\rangle$ $(|a|,|b| \neq
\{0,1\})$ the particular sequence of bit values 0 and 1 obtained
cannot, not even in principle\footnote{As theorems like those of
Kochen-Specker \cite{kochen} show, it is fundamentally not
possible to assign to a quantum system (noncontextual) properties
corresponding to all possible measurements. The theorems assert
that for a quantum system described in a Hilbert space of
dimension equal to or larger than three, it is possible to find a
set of $n$ projection operators which represent the yes-no
questions about an individual system, such that none of the $2^n$
possible sets of answers is compatible with the sum rule of
quantum mechanics for orthogonal decomposition of identity
\cite{peres} (i.e. if the sum of a subset of mutually commuting
projection operators is the identity one and only one of the
corresponding answers ought to be ''yes''). This means that it is
not possible to assign a definite unique answer to every single
yes-no question represented by a projection operator independent
of which subset of mutually commuting projection operators one
might consider it with together. If there are no definite
(context-independent) answers to all possible yes-no questions
that can be asked about the system then the operational concept of
the Shannon measure of information itself, defined as the number
of yes-no questions needed to determine the particular answers the
system gives, becomes highly problematic.}, be considered in any
way to be defined before the measurements are performed. The
non-existence of well-defined bit values prior to and independent
of observation suggests that the Shannon measure, as defined by
the number of binary questions needed to determine the particular
{\it observed} sequence 0's and 1's, becomes problematic and even
untenable in defining our uncertainty as given {\it before} the
measurements are performed.

Here we will critically analyze the applicability of the axiomatic
derivation of the Shannon measure for the case of quantum
measurement. We will also show that Shannon information is not
useful in defining the information content in a quantum system. In
fact we will see that when we try to apply Shannon's postulate in
quantum measurements or when we try to define the information
content by the Shannon information a certain element emerges that
escapes complete and full description in quantum mechanics. This
element is always associated with the objective randomness of
individual quantum events and with quantum complementarity. In the
end we will briefly discuss a novel and more suitable measure of
information \cite{caslav}. Yet at first we will return to a
discussion in more detail of the operational definition of Shannon
information to quantum measurements.

\section{Discussion of the Operational Definition for a Sequence of Measurements}

For classical observations Shannon's measure of information can
conceptually be motivated through an operational approach to the
question. We will follow the introduction of Shannon's measure of
information as given by Uffink \cite{uffink}. Consider an urn
filled with N colored balls. There are $n_1$, $n_2$, ..., $n_m$
balls with various different colors: black, white, ..., red. Now
the urn is shaken, and we draw one after the other all balls from
the urn. To what extent can we predict the particular color
sequence drawn?

Certainly, if all the balls in the urn are of the same color, we
can completely predict the color sequence. On the other hand, if
the various colors are present in equal proportions and if we have
no knowledge about the arrangement of the balls after shaking the
urn, we are maximally uncertain about the color sequence drawn. As
noticed in \cite{uffink} one can think of these situations as
extreme cases on a varying scale of predictability. For example,
for N=4, there is only one color sequence
$\circ$$\circ$$\circ$$\circ$ if all balls are white, 4 possible
color sequences $\bullet$$\circ$$\circ$$\circ$,
$\circ$$\bullet$$\circ$$\circ$, $\circ$$\circ$$\bullet$$\circ$,
$\circ$$\circ$$\circ$$\bullet$, if there are three black and one
white ball in the urn, yet 6 possible color sequences
$\bullet$$\bullet$$\circ$$\circ$,
$\bullet$$\circ$$\bullet$$\circ$,
$\bullet$$\circ$$\circ$$\bullet$,
$\circ$$\bullet$$\bullet$$\circ$, $\circ$$\bullet$$\circ$$\bullet$
and $\circ$$\circ$$\bullet$$\bullet$ if there are two black and
two white balls in the urn. This suggest that the uncertainty we
have before drawing about the particular color sequence that will
be drawn is defined by the total number of different possible
color sequences that are in accordance with the given number of
balls with their respective colors in the urn.

Consider now a situation where a long sequence of N balls are
drawn from an infinite "sea" of balls with proportions $p_1$,
$p_2$, ..., $p_m$ for the different colours in the sea. Then a
long sequence contains with high probability about $p_1N$ balls of
the first colour, $p_2N$ balls of the second colour etc. (such a
sequence is called typical sequence). The probability to obtain a
particular typical sequence (particular colour sequence) is given
by \cite{shannon}
\begin{equation}
p(sequence) = p_1^{p_1N} p_2^{p_2N} ... p_m^{p_mN} =
\frac{1}{2^{NH}}
\end{equation}
where
\begin{equation}
H=-\sum_{i=1}^{m} p_i \log p_i \label{who}
\end{equation}
is the Shannon information expressed in bits with the logarithm
taken to base 2. Consequently, the total number of distinct
typical sequences is given by
\begin{equation}
W \simeq 2^{N H}. \label{christ}
\end{equation}


Suppose now that one wishes to identify a specific color sequence
of the drawn balls from the complete set of possible color
sequences by asking questions to which only ''yes'' or ''no'' can
be given as an answer. Of course, the number of questions needed
will depend on the questioning strategy adopted. In order to make
this strategy the most optimal, that is, in order that we can
expect to gain maximal information from each yes-or-no question,
we evidently have to ask questions whose answers will strike out
always half of the possibilities.

\begin{figure}
\centerline{\psfig{width=9.5cm,file=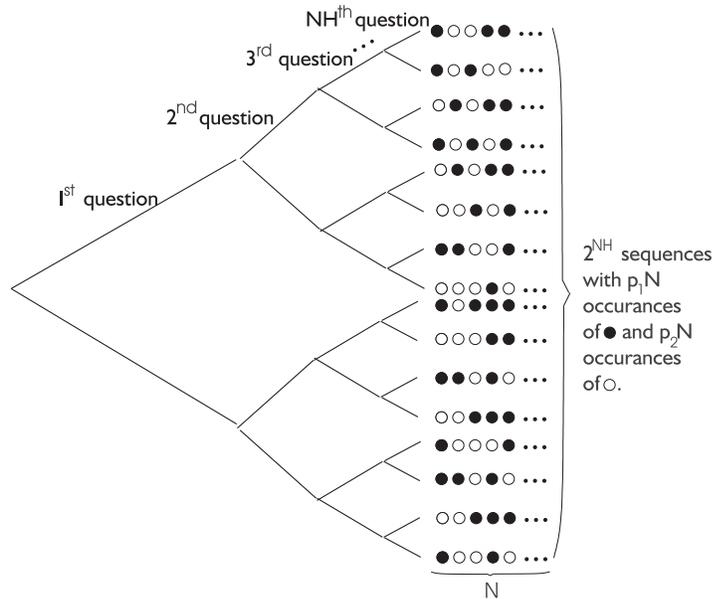}} \caption{Binary
question tree to determine the specific sequence of outcomes
(color of the drawn balls) in a sufficiently large number $N$ of
experimental trials (number of drawings). An urn is filled with
black and white balls with proportions $p_1$ and $p_2$,
respectively. The expected number of questions needed to determine
the actual (typical) sequence of outcomes is $NH$, where
$H=-p_1\log p_1 -p_2\log p_2$.} \label{generalbaum}
\end{figure}

Since there are $W=2^{NH}$ possible different (typical) color
sequences (all of them have equal probability to be drawn), the
minimal number of yes-no questions needed is just $NH$. Or
equivalently, the Shannon information expressed in bits is the
minimal number of yes-no questions necessary to determine which
particular sequence of outcomes occurs, divided by $N$
\cite{feinstein,uffink,blabla}. A particular color sequence is
specified by writing down, in order, the yes's and no's
encountered in traveling from the root to the specific leaf of the
tree as schematically depicted in Fig. \ref{generalbaum} for an
explicit example with an urn containing black and white balls
only.

If instead of balls with pre-assigned colors we consider quantum
systems whose individual properties are not defined before the
measurements are performed, does the Shannon measure of
information still define the information gain in the measurements
appropriately? More precisely, we ask here the question whether
the total number $W=2^{NH}$ of different possible (typical)
sequences of outcomes is suitable as a measure of our uncertainty
before the sequence of quantum measurements is performed.

In classical physics the behavior of the whole ensemble follows
from the behavior of its intrinsic different individual
constituents which can be thought of as being defined to any
precision. This is not the case in quantum mechanics. The
principal indefiniteness, in the sense of fundamental nonexistence
of a detailed description of and prediction for the individual
quantum event resulting in the particular measurement result,
implies that the particular sequence of outcomes specified by
writing down, in order, the yes's and no's encountered in a row of
yes/no questions asked is not defined before the measurements are
performed. No definite outcomes exists before measurements are
performed and therefore the number of different possible sequence
of outcomes does not characterize our uncertainty about the
individual system given before measurements are performed.

However, once the sequence of quantum measurements is performed
and the measurement results are obtained, the measure of
information needed to specify the particular sequence of outcomes
realized is defined appropriately by the Shannon measure. In the
sense that an individual quantum event manifests itself only in
the measurement process and is not precisely defined before
measurement is performed, we may speak of ''generation'' of that
specific information in the measurement.

\section{Inapplicability of Shannon's Postulates in Quantum Measurements}
\label{zika}

As observed by Uffink \cite{uffink}, an important reason for
preferring the Shannon measure of information lies in the fact
that it is uniquely characterized by Shannon's intuitively
reasonable postulates. This has been expressed strongly by Jaynes
\cite{jaynes} : ''One ... important reason for preferring the
Shannon measure is that it is the only one that satisfies ...
[Shannon's postulates]. Therefore one expects that any deduction
made from other information measures, if carried far enough, will
eventually lead to contradiction.'' A good way to continue our
discussion is by reviewing how Shannon, using his postulates,
arrived at his famous expression. He writes \cite{shannon}:

''Suppose we have a set of possible events whose probabilities of
occurrence are $p_1,p_2,...,p_n$. These probabilities are known
but that is all we know concerning which event will occur. Can we
find a measure of how much ''choice'' is involved in the selection
of the event or how uncertain we are of the outcome?

\begin{figure}
\centerline{\psfig{width=6.7cm,file=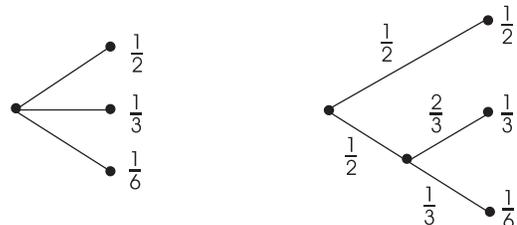}}
\caption{Decomposition of a choice from three possibilities.
Figure taken from [1].} \label{choicefig}
\end{figure}

If there is such a  measure, say $H(p_1,p_2,...,p_n)$, it is
reasonable to require of it the following properties:
\begin{enumerate}
\item $H$ should be continuous in the $p_i$.
\item If all the $p_i$ are equal, $p_i=\frac{1}{n}$, then $H$ should
be a monotonically increasing function of $n$. With equally likely
events there is more choice, or uncertainty, when there are more
possible events.
\item If a choice be broken down into two successive choices, the
original $H$ should be the weighted sum of the individual values
of $H$. The meaning of this is illustrated in Fig.
\ref{choicefig}. At the left we have three possibilities
$p_1=\frac{1}{2}$, $p_2=\frac{1}{3}$, $p_3=\frac{1}{6}$. On the
right we first choose between two possibilities each with
probability $\frac{1}{2}$, and if the second occurs make another
choice with probabilities $\frac{2}{3}$, $\frac{1}{3}$. The final
results have the same probabilities as before. We require, in this
special case, that
\[
H\left(\frac{1}{2},\frac{1}{3},\frac{1}{6}\right) =
H\left(\frac{1}{2},\frac{1}{2}\right) + \frac{1}{2}
H\left(\frac{2}{3},\frac{1}{3}\right).
\]
The coefficient $\frac{1}{2}$ is the weighing factor introduced
because this second choice occurs half the time.''
\end{enumerate}

Shannon then shows that only the function (\ref{who}) satisfies
all three postulates. It is the third postulate which determines
the logarithm form of the function and, as we will argue, it is
this postulate which leads to problems when quantum measurements
are involved.

We now turn to the discussion of Shannon's postulates. While the
first two postulates are natural for every meaningful measure of
information, the last postulate might deserve more justification
The third Shannon postulate originally formulated as an example
was reformulated as an exact rule by Faddeev
\cite{faddeev,uffink}: For every $n\geq 2$
\begin{equation}
H(p_1,..,p_{n-1},q_1,q_2)\!= \!H(p_1,..,p_{n-1},p_n)\! +\! p_n
H\!\left(\frac{q_1}{p_n},\frac{q_2}{p_n}\right), \label{popmilo}
\end{equation}
where $p_n=q_1+q_2$.

Without physical interpretation the recursion postulate (the name
was suggested in \cite{uffink}) (\ref{popmilo}) is merely a
mathematical expression which is certainly necessary for the
uniqueness of the function (\ref{who}) but has no further physical
significance. We adopt the following well-known interpretation
\cite{uffink,jaynes0.5}. Assume the possible outcomes of the
experiment to be $a_1,...,a_n$ and $H(p_1,...,p_n)$ to represent
the amount of information that is gained by the performance of the
experiment. Now, decompose event $a_n$ into two distinct events
$a_n \wedge b_1$ and $a_n \wedge b_2$ (''$\wedge$'' denotes
''and'', thus $a \wedge b$ denotes a joint event). Denote the
probabilities of outcomes $a_n \wedge b_1$ and $a_n \wedge b_2$ by
$q_1$ and $q_2$, respectively. Then the left-hand side
$H(p_1,...,p_{n-1},q_1,q_2)$ of Eq. (\ref{popmilo}) represents the
amount of information that is gained by the performance of the
experiment with outcomes $a_1,...,a_{n-1},a_n \wedge b_1,a_n
\wedge b_2 $.

When the outcome $a_n$ occurs, the conditional probabilities for
$b_1$ and $b_2$ are $\frac{q_1}{p_n}$ and $\frac{q_2}{p_n}$
respectively and the amount of information gained by the
performance of the conditional experiment is
$H\left(\frac{q_1}{p_n},\frac{q_2}{p_n}\right)$. Hence the
recursion requirement states that the information gained in the
experiment with outcomes $a_1,...,a_{n-1}, a_n \wedge b_1, a_n
\wedge b_2 $ equals the {\it sum} of the information gained in the
experiment with outcomes $a_1,...,a_n$ and the information gained
in the conditional experiment with outcomes $b_1$ or $b_2 $, given
that the outcome $a_n$ occurred with probability $p_n$.

This interpretation implies that the third postulate can be
rewritten as
\begin{eqnarray}
&H&(p(a_1),...,p(a_{n-1}),p(a_n \wedge b_1),p(a_n \wedge b_2))
\label{popciro} \\ &=& H(p(a_1),...,p(a_{n-1}),p(a_n)) +
p(a_n)H(p(b_1|a_n),p(b_2|a_n))\nonumber
\end{eqnarray}
where
\begin{eqnarray}
&p&(a_n) = p(a_n \wedge b_1) + p(a_n \wedge b_2)\nonumber  \\
&p&(a_n\wedge b_1)= p(a_n) p(b_1|a_n) \mbox{ and } \\ &p&(a_n
\wedge b_2) = p(a_n) p(b_2|a_n). \nonumber
\end{eqnarray}
Here $p(b_i|a_n ) $ $i=1,2$ denotes the conditional probability
for outcome $a_n$ given the outcome $b_i$ occurred and $ p(a_n
\wedge b_i)$ denotes the joint probability that outcome $a_n
\wedge b_i$ occurs.

If we analyze the generalized situation with $n$ outcomes $a_i$ of
the first experiment $A$, $m$ outcomes $b_j$ of the conditional
experiment $B$ and $m n$ outcomes $a_i \wedge b_j$ of the joint
experiment $A \wedge B$, we may then rewrite the recursion
postulate in a short form as
\begin{equation}
H(A \wedge B) = H(A) + H(B|A) \label{hannah}
\end{equation}
where $H(B|A) = \sum_j^{n} p(a_j) H(b_1|a_j,...,b_m|a_j)$ is the
average information gained by observation $B$ given that the
conditional outcome $a_j$ occurred weighted by probability
$p(a_j)$ for $a_j$ to occur.

It is essential to note that the recursion postulate is inevitably
related to the manner in which we gain information in a classical
measurement. In fact, in classical measurements it is always
possible to assign to a system simultaneously attributes
corresponding to all possible measurements, here $a_i$, $b_j$ and
$a_i \wedge b_j$. Also, the interaction between measuring
apparatus and classical system can be thought to be made
arbitrarily small so that the experimental determination of $A$
has no influence on our possibility to predict the outcomes of the
possible future experiment $B$. In conclusion, the information
expected in a classical experiment from the joint experiment $A
\wedge B$ is simply the sum of the information expected from the
first experiment $A$ and the conditional information of the second
experiment $B$ with respect to the first, as expressed in Eq.
(\ref{hannah}).

Therefore, only for the special case of commuting, i.e.
simultaneously definite observables, the axiomatic derivation of
the Shannon measure of information is applicable and the use of
the Shannon information is justified to define the uncertainty
given before quantum measurements are performed. However, in
general, if $A$ and $B$ are noncommuting observables, the joint
probabilities on the left-hand side of Eq. (\ref{popciro}) cannot
in principle be assigned to a system simultaneously, and
consequently Shannon's crucial third postulate which is necessary
for the uniqueness of Shannon's measure of information ceases to
be well-defined.

Having seen that the third Shannon postulate in general is not
applicable in quantum measurements we next introduce two
requirements that are immediate consequences of Shannon's
postulates and in which all the probabilities that appear are
well-defined in quantum mechanics. We will show that the two
requirements are violated by the information gained in quantum
measurements implying that the Shannon measure loses its
preferential status with respect to alternative expressions when
applied to define information gain in quantum measurements.

\begin{figure}
\centerline{\psfig{width=5cm,file=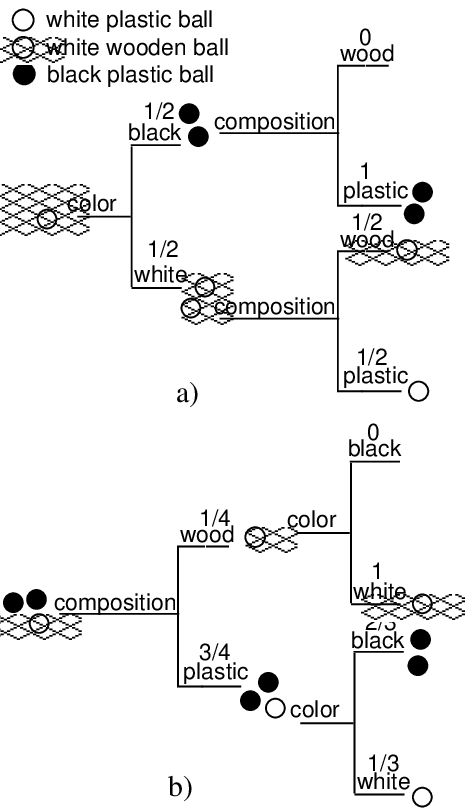}}
\caption{Indifference of information to the order of its
acquisition in classical measurements. A box is filled with balls
of different compositions (plastic and wooden balls) and different
colors (black and white balls). Now, the box is shaken. In (a) we
first draw a ball asking about the color of the drawn ball and
gain $H(\mbox{color})=1$ bit of information. Subsequently, we put
the black and white balls in separate boxes, draw a ball from each
box separately and ask about the composition of the drawn ball. We
gain $H_{bl}(\mbox{comp.})=0$ bits for the black balls and
$H_{wh}(\mbox{comp.})=1$ bit for the white balls. In (b) we pose
the two questions in the opposite order. We firstly ask about the
composition of the drawn ball and gain $H(\mbox{comp.})=0.81$
bits. In a conditional drawing we ask about the color of the drawn
ball and gain $H_{wo}(\mbox{color})=0$ bits for wooden balls and
$H_{pl}(\mbox{color})=0.92$ bits for plastic balls. The total
information gained is independent of the particular order the two
questions are posed, i.e. $H(\mbox{color})+ 1/2
H_{bl}(\mbox{comp.}) +1/2 H_{wh}(\mbox{comp.})$
$=\!H(\mbox{comp.})\!+\!1/4 H_{wo}(\mbox{color}) \!+\!3/4
H_{pl}(\mbox{color})\!=1.5$.} \label{balls1}
\end{figure}

\begin{enumerate}
\item {\it Every new observation reduces our ignorance and increases
our knowledge.} In his work Shannon \cite{shannon} offers a list
of properties to substantiate that $H$ is a reasonable measure of
information. He writes: ''It is easily shown that
\[
H(A\wedge B) \leq H(A) + H(B)
\]
with equality only if the events are independent (i.e., $p(a_i
\wedge b_j)=p(a_i)p(b_j))$. The uncertainty of a joint event is
less than or equal to the sum of the individual uncertainties''.
He continues further in the text: ''... we have
\[
H(A) + H(B) \geq H(A \wedge B)= H(A) + H(B|A).
\]
Hence,
\begin{equation}
H(B) \geq H(B|A). \label{mrgo}
\end{equation}
The uncertainty of $B$ is never increased by knowledge of $A$. It
will be decreased unless $A$ and $B$ are independent events, in
which case it is not changed'' (we have changed Shannon's notation
to coincide with that of our work).

\item {\it Information is indifferent on the order of acquisition.}
The total amount of information gained in successive measurements
is independent of the order in which it is acquired, so that the
amount of information gained by the observation of $A$ followed by
the observation of $B$ is equivalent to the amount of information
gained from the observation of $B$ followed by the observation of
$A$
\begin{equation}
H(A) + H(B|A) = H(B) + H(A|B). \label{blondie}
\end{equation}
This is an immediate consequence of the recursive postulate which
can be obtained when we write the recursion postulate in two
different ways depending on whether the observation of $A$ is
followed by the observation of $B$ or vice versa. An explicit
example for a sequence of classical measurements is given in Fig.
\ref{balls1}.
\end{enumerate}

\begin{figure}
\centerline{\psfig{width=7.8cm,file=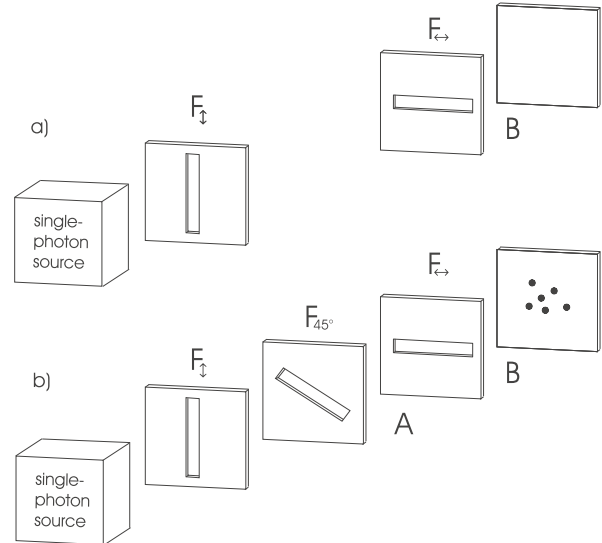}} \caption{The
gain of knowledge in a new observation reduces our knowledge at
hand from a previous observation. Filters $F_{\updownarrow}$,
$F_{45^\circ}$ and $F_{\leftrightarrow}$ are oriented vertically,
at $+45^{\circ}$ and horizontally, respectively. If filter
$F_{\leftrightarrow}$ is inserted behind the filter
$F_{\updownarrow}$, no photons are observed at the detector plate
(Fig. a). In this case our knowledge about horizontal polarization
(property $B$) of a photon passing through filter
$F_{\updownarrow}$ is complete. If filter $F_{45^\circ}$ is
inserted between $F_{\updownarrow}$ and $F_{\leftrightarrow}$, a
certain number of photons (1/4 of the number of photon passing
through $F_{\updownarrow}$) will be observed at the detection
plate (Fig. b). Now acquisition of information about the
polarization at $+ 45^{\circ}$ (property $A$) leads to the
decrease of our previous knowledge about horizontal polarization
of the photon.} \label{filters}
\end{figure}

Are these two requirements satisfied by information gained in
quantum measurements? Consider a beam of randomly polarized
photons. Filters $F_{\updownarrow}$, $F_{45^\circ}$ and
$F_{\leftrightarrow}$ are oriented vertically, at $+45^{\circ}$,
and horizontally respectively, and can be placed so as to
intersect the beam of photons (Fig. \ref{filters}). If we insert
filter $F_{\updownarrow}$ the number of photons observed at the
detection plate will be approximately half of the number in the
incoming beam. The outgoing photons now all have vertical
polarization. Notice that the function of filter
$F_{\updownarrow}$ cannot be explained as a ''sieve'' that only
lets those photons pass that are {\it already} of vertical
polarization in the incoming beam. If that were the case, only a
certain small number of the randomly polarized incoming photons
would have vertical polarization, so we would expect a much larger
attenuation of the beam of photons as they pass the filter.

Denote with $A$ and $B$ properties of the photon to have
polarization at $+45^{\circ}$ and horizontal polarization,
respectively. If $F_{\leftrightarrow}$ is inserted behind the
filter $F_{\updownarrow}$ we are certain that none of the photons
will pass through (Fig. \ref{filters}a). For a photon with
vertical polarization we have complete knowledge of the property
$B$, i.e. $H(B)=0$. Notice that a ''sieve'' model could explain
this behaviour. If we now insert $F_{45^\circ}$ between
$F_{\updownarrow}$ and $F_{\leftrightarrow}$ we observe an effect
which cannot be explained by a sieve model where the filter does
not change the object. However we now observe a certain number of
photons at the detection plate (about $\frac{1}{4}$ of the number
of photons in the beam passed through $F_{\updownarrow}$) as shown
in Fig. \ref{filters}b. In this case our knowledge of the property
$B$ is not complete anymore.

\begin{figure}
\centerline{\psfig{width=8.1 cm,file=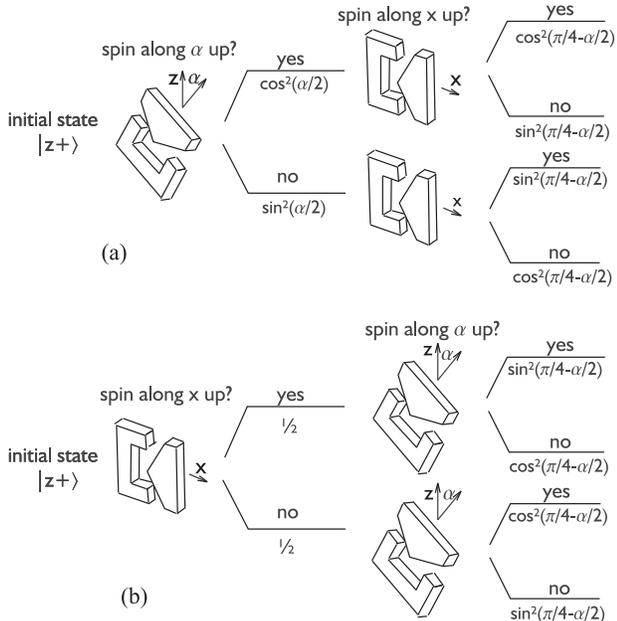}}
\caption{Dependence of information on the order of its acquisition
in successive quantum measurements. A spin-1/2 particle is in the
state $|z+\rangle$ spin-up along the z-axis. Spin along the x-axis
and spin along the direction in the x-z plane tilted at an angle
$\alpha$ from the z-axes are successively measured, in one order
in (a) and in the opposite order in (b). Whereas we obtain an
equal portion $H(\cos^2(\pi/4-\alpha/2), \sin^2(\pi/4-\alpha/2))$
of information in the conditional (subsequent) measurement both in
(a) and in (b), the amounts of information
$H(\cos^2\alpha/2,\sin^2\alpha/2)$ and
$H(\frac{1}{2},\frac{1}{2})=1$ we gain in the first measurement in
(a) and in the first measurement in (b) respectively, can be
significantly different. Specifically for $\alpha \rightarrow 0$
we have complete knowledge about spin along the direction at the
angle $\alpha$ in (a) but absolutely no knowledge about the spin
along the x-axis in (b). We emphasize that we do not assume any
specific functional dependence for the measure of information
$H$.} \label{qballs}
\end{figure}

The acquisition of information about property $A$ therefore leads
to a decrease of our knowledge about property $B$, i.e. $H(B|A) >
0$. Note that on the photons absorbed by the filter $F_{45^\circ}$
we cannot measure property $B$ subsequently. However already for
the subensemble of the photons passing through the filter our
uncertainty about property $B$ becomes larger than 0 implying
$0=H(B)< H(B|A)$ which clearly violates requirement (\ref{mrgo}).
Another example of sequence of quantum measurements where
requirement (\ref{blondie}) is violated is given in Fig.
\ref{qballs}. Clearly, violation of the requirements (\ref{mrgo})
and (\ref{blondie}) occurs when the corresponding operators $A$
and $B$ do not commute.

What is the origin of the violation of the requirements
(\ref{mrgo}) and (\ref{blondie}) in quantum measurements? In
contrast to a classical measurement which just adds some new
knowledge to our knowledge at hand from the previous measurements,
in a quantum measurement the gain of the new knowledge is always
at the expense of irrecoverable loss of complementary classes of
knowledge. This originates from the distinction between ''total''
and ''complete'' information in quantum physics. In classical
physics the total information about a system is complete. In
quantum physics the total information of a system, represented by
the state vector, is never complete in the sense that all possible
future measurement results are precisely defined\footnote{Yet, we
do not hesitate to emphasize that it certainly is complete in the
sense that it is not possible to have more information about a
system than what can be specified in its quantum state. In fact,
the state vector represents that part of our knowledge about the
history of a system which is necessary to arrive at the maximum
possible set of probabilistic predictions for all possible future
observations of the system. For example, a set of complex
amplitudes of a $\psi$-function is a specific representation of
the catalog of our knowledge of the system. This view was assumed
by Schr\"{o}dinger \cite{schroedinger} who wrote: ''Sie ((die
$\psi$-Funktion )) ist jetzt das Instrument zur Vorausage der
Wahrscheinlichkeit von Ma{\ss}zahlen. In ihr ist die jeweils erreichte
Summe theoretisch begr\"{u}ndeter Zukunfterwartungen verk\"{o}rpert,
gleichsam wie in einem {\em Katalog} niedergelegt. Translated:
''It (the $\psi$-function) is now the means for predicting the
probability of measurement results. In it is embodied the
momentarily attained sum of theoretically based future
expectation, somewhat as laid down in a {\em catalog.''}}. In
fact, the total information of a quantum system suffices to
specify the eigenstate of one nondegenerate (with one-dimensional
eigenspaces only) observable only.

For example, the state of a photon passing through filter
$F_{\updownarrow}$ is specified by the complete knowledge about
the property $A$ of vertical polarization. If we let a photon in
this state pass through filter $F_{45^\circ}$ as given in Fig.
\ref{filters}b, our knowledge of the photon changes, and therefore
its representation, the quantum state, also changes. The total
information of a photon in the new state is completely exhausted
in specifying property $B$ of polarization at $45^\circ$ and no
further information is left to also specify property $A$, thus
implying unavoidable loss of the previous knowledge about this
property. This further implies that the set of future
probabilistic predictions specified by the new projected state is
indifferent to the knowledge collected from the previous
measurements in the whole history of the system. Such a view was
assumed by Pauli \cite{pauli} who writes\footnote{In translation:
''In the case of indefiniteness of a property of a system for a
certain experimental arrangement (for a certain state of the
system) any attempt to measure that property destroys (at least
partially) the influence of earlier knowledge of the system on
(possibly statistical) statements about later possible measurement
results.''}: ''Bei Unbestimmtheit einer Eigenschaft eines Systems
bei einer bestimmten Anordnung (bei einem bestimmten Zustand des
Systems) vernichtet jeder Versuch, die betreffende Eigenschaft zu
messen, (mindestend teilweise) den Einflu{\ss} der fr\"{u}heren Kenntnisse
vom System auf die (eventuell statistischen) Aussagen \"{u}ber sp\"{a}tere
m\"{o}gliche Messungsergebnisse.'' This clearly makes possible to
violate requirements (\ref{mrgo}) and (\ref{blondie}) in quantum
measurements.

Here a certain misconception might be put forward that arises from
a certain practical point of view. According to that view, for
example, complementarity between interference pattern and
information about the path of the particle in the double-slit
experiment is considered to arise from the fact that any attempt
to observe the particle path would be associated with an
uncontrollable disturbance of the particle. Such a disturbance in
itself would then be the reason for the loss of the interference
pattern. In such of view it would be possible to define Shannon's
information for all attributes of the system simultaneously, and
the third Shannon postulate, as well as the requirements
(\ref{mrgo}) and (\ref{blondie}), would be violated because of the
unavoidable disturbance of the system occurring whenever the
subsequently measured property $B$ is incompatible with the
previous one $A$. Yet, this is a misconception for two reasons.

Firstly, as theorems like those of Bell \cite{bell} or
Greenberger, Horne and Zeilinger \cite{ghz} show, it is not
possible, not even {\it in principle}, to assign to a quantum
system simultaneously observation-independent properties which in
order to be in agreement with special relativity have to be local.
We therefore cannot speak of a ''disturbance'' in the measurement
process if there are no objective properties to disturb.

Secondly, over the last few years experiments were considered and
some already performed, where the reason why no interference
pattern arises is not due to any uncontrollable disturbance of the
quantum system or the clumsiness of the apparatus. Rather the lack
of interference is due to the fact that the quantum state is
prepared in such a way as to permit path {\em information} to be
obtained, in principle, independent of whether the experimenter
cares to read it out or not. One line of such research considers
the use of micromasers in atomic beam experiments \cite{scully},
another one concerns experiments on correlated photon states
emerging from nonlinear crystals through the process of
parametric-down conversion \cite{horne}.

The view that complementarity must be based on the much more
fundamentally property of mutual exclusiveness of different
classes of information of a quantum system was assumed by Pauli
\cite{pauli} in the analysis of the uncertainty
relations\footnote{In translation:''... these relations contain
the statement that any precise knowledge of the position of a
particle implies a fundamental indefiniteness, not just an
unknownness, of the momentum for a consequence and vice versa. The
distinction between (fundamental) {\em indefiniteness} and {\em
unknownness}, and the relation between these two notions is
decisive for the whole quantum theory.''}: '' ... diese Relationen
enthalten die Aussage, da{\ss} jede genaue Kenntnis des Teilchenortes
zugleich eine prinzipielle Unbestimmtheit, nicht nur Unbekanntheit
des Impulses zur Folge hat und umgekehrt. Die Unterscheidung
zwischen (prinzipieller) {\it Unbestimmtheit} und {\it
Unbekanntheit} und der Zusammenhang beider Begriffe sind f\"{u}r die
ganze Quantentheorie entscheidend.''

\section{Difficulties in Defining the Information Content
of a Quantum System}

To define the information content of a physical system one might
consider different measures of information. However only those
measures of information have physical significance according to
which the defined information content of the system possesses
properties which naturally follow from the physical situation
considered. These properties are, for example, invariance under
changes of the modes of observation of the system and conservation
in time if there is no information exchange with an environment.
We show now that the information content of a quantum system, if
it is assumed to be measured by the Shannon measure of
information, cannot be defined in any way to have these
properties.

The classical world appears to be composed of particles and
fields, and the properties of each one of these constituents could
be specified quite independently of the particular phenomenon
discussed or of the experimental procedure a physicist chooses to
determine these properties. In other words the properties of
constituents of the classical world are noncontextual.

In particular, the total lack of information about a classical
pointlike system (with no rotational and internal degrees of
freedom) defined as Shannon's information associated with the
probability distribution over the phase space is independent of
the specific set of variables chosen to describe the system
completely (such as position and momentum, or bijective functions
of them) and conserved in time if there is no information exchange
with an environment (i.e. if the system is dynamically independent
from the environment and not exposed to a
measurement)\footnote{Given the probability distribution
$\rho(\vec{r},\vec{p},t) $ over the phase space the total lack of
information of a classical system is defined as \cite{jaynes1}
\begin{equation}
H_{total}(t) =-\int d^3 \vec{r} d^3 \vec{p}
\rho(\vec{r},\vec{p},t) \log \frac{\rho(\vec{r},\vec{p},t)}
{\mu(\vec{r},\vec{p})}, \label{dupe}
\end{equation}
where a background measure $\mu(\vec{r},\vec{p})$ is an additional
ingredient that has to be added to the formalism to ensure
invariance under change of variables when we consider continuous
probability distributions. The conservation of $H_{total}$ in time
for a system with no information exchange with an environment is
implied by the Hamiltonian evolution of a point in phase space.}.
Operationally the total information content of a classical system
can be obtained in the joint measurement of position and momentum,
or in successive measurements in which the observation of position
is followed by the observation of momentum or vice
versa\footnote{In full analogy with (\ref{blondie}) we may write
$H_{total}(\vec{r},\vec{p}) = H(\vec{r}) +  H(\vec{p}|\vec{r})=
H(\vec{p}) + H(\vec{r}|\vec{p})$.}.

Contrary to the classical concepts most quantum-mechanical
concepts are limited to the description of phenomena within some
well-defined experimental context, that is, always restricted to a
specific experimental procedure the physicist chooses. In
particular the amount of information gained in an individual
quantum measurement depends strongly on the specific experimental
context. In the optimal experiment when the measurement basis
$|i\rangle$ coincides with the eigenbasis of the density matrix
$\hat{\rho}$ of the system: $\hat{\rho} |i\rangle=w_i|i\rangle$
the amount of information gained is maximal (See for example
\cite{peres}). Since in the basis corresponding to the optimal
experiment the density operator is represented by a diagonal
matrix with elements $w_i$, the information gain defined by the
Shannon measure equals the von Neumann entropy as given
by\footnote{For a {\em given} density matrix $\hat{\rho}$ the von
Neumann entropy
\begin{equation}
S(\hat{\rho}) = -Tr(\hat{\rho} \log \hat{\rho})
\end{equation}
is widely accepted as a suitable definition for the information
content of a quantum system. For a system described in
$N$-dimensional Hilbert space this ranges from $\log N$ for a
completely mixed state to 0 for a pure state. The von Neumann
entropy has the important property to be invariant under unitary
transformations. However, we observe that any function of the form
$Tr(f(\hat{\rho}))$ (the operator $f(\hat \rho)$ is identified by
having the same eigenstates as $\hat{\rho}$ and the eigenvalues
$f(w_j)$, equal to the function values taken at the eigenvalues
$w_j$ of $\hat{\rho}$.) possesses this invariance property. We
also observe that the von Neumann entropy is a property of the
quantum state as a whole without explicit reference to information
contained in individual measurements.}
\begin{equation}
H = -\sum_i w_i \log w_i = -Tr(\hat{\rho} \log \hat{\rho}).
\label{klasik}
\end{equation}
This has the important property to be invariant under unitary
transformations $\hat{\rho} \rightarrow \hat{U}\hat{\rho}
\hat{U}^+$. The invariance under unitary transformations implies
invariance under the change of the representation (basis) of
$\hat{\rho}$ and conservation in time if there is no information
exchange with an environment. The later precisely means that if we
perform the optimal experiments both at time $t_0$ and at some
future time $t$, Shannon's information measures associated to the
optimal experiments at the two times will be the same, i.e.
\begin{equation}
H(t)= -\sum_i w_i(t) \log w_i(t) = -\sum_i w_i \log w_i = H(t_0).
\end{equation}
Here, the eigenvalues of the density matrix at time $t$ are
$w_i(t)=w_i$.

However, without the {\it additional} knowledge of the eigenbasis
of the density matrix $\hat{\rho}$ we cannot find the optimal
experiment and obtain directly the Shannon information associated.
Also, all the statistical predictions that can be made for the
optimal measurement are the same as if we had an ordinary
(classical) mixture, with fractions $w_i$ of the systems giving
with certainty results that are associated to the eigenvectors
$|i\rangle$. In this sense the optimal measurement is a classical
type measurement and therefore in this particular case, and only
then, Shannon's measure defines the information gain in a
measurement appropriately\footnote{Consider a situation where
instead using of single systems to send information to the
receiver a sender uses a sequence of $N$ systems where each
individual system is drawn from an ensemble of pure states
$\{|\psi_1\rangle, ..., |\psi_n\rangle\}$, with frequency of
occurrence $ \{w_1,...,w_n\}$ respectively. It was shown in
\cite{hausladen} that for sufficiently large $N$ there are $2^{N
S(\hat{\rho})}$ highly distinguishable sequences of pure states
which become mutually orthogonal as $N \rightarrow \infty$. Here
$S(\hat{\rho})=-Tr(\hat{\rho} \log \hat{\rho})$ is the von Neumann
entropy and $\hat{\rho} = \sum^n_i w_i |\psi_i\rangle \langle
\psi_i|$. This means that if the sender uses a sequence consisting
of a choice of states that respects the a priori frequencies
$w_i$, and the receiver distinguishes whole sequences rather than
individual states, then the (Shannon) information transmitted per
system can be made arbitrary close to $S(\hat{\rho})$. Here again
the total density matrix $\hat{\rho}^N$ of $N$ systems can be made
arbitrary close to the one as if we had a classical mixture of the
$2^{N S(\hat{\rho})}$ sequences of states.}. Considering also our
previous discussion it is therefore not surprising that Shannon's
measure is useful only when applied to measurements which can be
understood as classical measurements.

Which set of individual measurements should we perform and how to
combine individual measures of information obtained in the set in
order to arrive at the information content of a quantum system if
we do not know the eigenbasis of the density matrix? Quantum
complementarity implies that the total information content of the
system might be partially encoded in different mutually exclusive
(complementary) observables. These have the property that complete
knowledge of the eigenvalue of any one of the observables excludes
{\it any} knowledge about the eigenvalues of all other
observables. Such a set of observables for a spin-1/2 particle can
for example be spin components along orthogonal directions.

We consider now a quantum system described in $n$ dimensional
Hilbert space and we denote a complete set of $m$ mutually
complementary observables\footnote{To specify a system described
by a $n \times n$ density matrix completely one needs $n^2-1$
independent real numbers. Any individual, complete measurement (we
consider here only complete measurements, i.e., where operators
associated to the measurements are without degeneracy) with $n$
possible outcomes defines $n-1$ independent probability values
(the sum of all probabilities for all possible outcomes in an
individual experiment is one). Therefore, just on the basis of
counting the number of independent variables, we expect that the
number of different measurements we need in order to determine the
density matrix completely is $\frac{n^2-1}{n-1}=n+1$. Ivanovic
\cite{ivanovic}, and Wootters and Fields \cite{wootters}
demonstrated the existence of exactly $n+1$ mutually complementary
observables by an explicit construction in the cases of $n$ prime
and $n=2^k$.} by $\{\hat{A}, \hat{B}, ...\}$. The property of
mutual expansiveness implies that if the system is in an
eigenstate of one of the observables, for example, in the
eigenstate $|a_j\rangle$ of the observable $\hat{A}$ and we
measure any other observable from the set, say $\hat{B}$,
projecting the system onto states
$\{|b_1\rangle,...,|b_i\rangle,...,|b_n\rangle\}$, the individual
outcome is completely random (all measurement results are equally
probable)
\begin{equation} |\langle
a_j|b_i\rangle|^2 = \frac{1}{n} \hspace{0.7cm} \forall i,j.
\label{goja}
\end{equation}

It was shown in \cite{ivanovic} that the density matrix of the
system can fully be reconstructed if one performs a complete set
of mutually complementary observations. This suggest that the
total information content of a quantum system represented by a
density matrix $\hat{\rho}$ is {\it all} obtainable from a
complete set of mutually complementary measurements. To obtain the
total information one however cannot perform the set of
measurements successively because, unlike the classical case, the
information obtained in successive quantum measurements depends on
the order of its acquisition (see Fig. \ref{qballs} and discussion
above). Instead it seems that any attempt to obtain the total
information content of a quantum system has to be related to the
complete set of mutually complementary experiments performed on
systems that are all in the same quantum state.

We suggest that it is therefore natural to require that the total
information content in a system in the case of quantum systems is
{\it sum} of the individual amounts of information over a complete
set of $m$ mutually complementary observables. As already
mentioned above, for a spin-1/2 particle these are three spin
projections along orthogonal directions. If we define the
information gain in an individual measurement by the Shannon
measure the total information encoded in the three spin components
is given by
\begin{equation}
H_{total}:=H_1(p^+_x,p^-_x)+H_2(p^+_y,p^-_y)+H_3(p^+_z,p^-_z).
\label{tic1}
\end{equation}
Here, e.g. $p^+_x$ is the probabilities to find the particle with
spin up along direction $x$.

Considering now an explicit example we will show that the total
information $H_{total}$ based on the Shannon measure is in general
{\it not} invariant under unitary transformations. We calculate
(\ref{tic1}) for a spin-1/2 particle in the state $|\psi \rangle =
\cos\theta/2 |z+\rangle + \sin\theta/2 |z-\rangle$ and we find
that
\begin{eqnarray}
&H&_{total} = \\ &-& \frac{1-\sin\theta}{2} \log
\frac{1-\sin\theta}{2} - \frac{1+\sin\theta}{2} \log
\frac{1+\sin\theta}{2}\nonumber \\  &-&\cos^2\!\frac{\theta}{2}
\log\left(\cos^2 \frac{\theta}{2}\right) -  \sin^2\frac{\theta}{2}
\log\left(\sin^2\!\frac{\theta}{2}\right) +1 \nonumber
\end{eqnarray}
depends on the parameter $\theta$, thus being not invariant under
unitary transformations. This associates a number of highly
counter-intuitive properties to $H_{total}$: 1) it can be
different for states of the same purity (e.g. it takes its maximal
value of 2 bits of information for $\theta=0$ and it takes its
minimal value of 1.36 bits for $\theta=\pi/4$); 2) it changes in
time even for a system completely isolated from the environment
where no information can be exchanged with environment; 3) it can
take different values for different sets of the three orthogonal
spin projections. These unnatural properties we see again as a
strong indications for inadequacy of the Shannon measure to define
the information gain in an individual quantum measurement.

\section{A Suggested Alternative Measure of Information}

We suggest that it is natural to require that the information
content of the quantum system defined as a sum of individual
measures over a complete set of mutually complementary
measurements is invariant under unitary transformations. Having
shown that this cannot be achieved with the Shannon measure of
information we now introduce a new measure of information that
differs both mathematically and conceptually from Shannon's
measure of information and according to which the information
content has invariance property.

The new measure of information is a quadratic function of
probabilities\footnote{Expressions of the general type of Eq.
(\ref{junko}) were studied in detail by Hardy, Littlewood and
P\'{o}lya \cite{hardy}. They introduced a general class of
mathematical expressions
\begin{equation}
M_\alpha = \left(\sum_{i=1}^{n} p^\alpha_i \right)^{\alpha-1}
\mbox{ for } 0 \leq \alpha \leq \infty
\end{equation}
that from the point of view of information theory all can be
assumed to quantify information properly. These expressions are
also closely related to Tsallis's \cite{tsallis} nonextensive
entropy $S_{\alpha}= \frac{1}{1-\alpha} \sum_{i=1}^{n} (p^\alpha_i
-1)$ and R\'{a}nyi's \cite{ranyi} entropy $H_{\alpha} =
\frac{1}{1-\alpha} \log \sum_{i=1}^{n} p^\alpha_i.$}
\begin{equation} I(p_1,...,p_n) = \sum_{i=1}^{n}
\left(p_i-\frac{1}{n}\right)^2, \label{junko}
\end{equation}
and it takes into account that for quantum systems the only
features known before an experiment is performed are the
probabilities for various events to occur (See \cite{caslav} for
discussion; there a specific normalization factor in expression
(\ref{junko}) was used resulting in maximally $k$ bits for $n=2^k$
possible outcomes). The measure $I(p_1,...,p_n)$ takes its maximal
value of $(n-1)/n$ if one $p_i=1$ and it takes its minimal value
of 0 when all $p_i$ are equal.

The important property of the new measure of information is that
the total information defined with respect to it is {\it
invariant} under unitary transformations. Using Eq. (\ref{goja})
one obtains that the sum over individual measures of information
of mutually complementary observations results in \cite{phd}
\begin{eqnarray}
I_{total}&:=& \sum_{j=1}^{m} I(p^j_1,...,p^j_n) \label{ljubav}
\\&=& \sum_{j=1}^{m} \sum_{i=1}^{n} \left(p^j_i -
\frac{1}{n}\right)^2 = Tr\hat{\rho}^2 -\frac{1}{n},\nonumber
\end{eqnarray}
for a system described by  the density matrix $\hat{\rho}$. Here
$p^j_i$ denotes the probability to observe the $i$-th outcome of
the $j$-th observable. The total information content of the system
therefore might all be encoded in one single observable or,
alternatively it might be partially encoded in all $m$ mutually
complementary observables. For a composite system in a product
state the total information can all be encoded in individual
systems constituting the composite system or, alternatively in the
extreme case of maximally entangled states it can all be encoded
in joint properties of the systems with no information left in
individual systems \cite{caslav}.

Independent of the various possibilities to encode information the
total information content of the system cannot fundamentally
exceed the maximal possible amount of information that can be
encoded in an individual observable $[=(n-1)/n]$. This upper limit
is reached when the system is in the pure state. When the system
is in a completely mixed state the total information takes its
minimal value of $0$.

The property of invariance under unitary transformations implies
that the total information content of the system does not
dependent of the particular set of mutually complementary
observables; it is a characterization of the state of the system
alone, not of the specific reference set of complementary
observables. Furthermore, since evolution in time is described by
a unitary operation the total information of the system is
conserved in time if there is no information exchange with the
environment.

We would like to note that the total information (\ref{ljubav})
was used in \cite{lee} to study the transfer of entanglement and
information for quantum teleportation of an unknown entangled
state through noisy quantum channels. The total information
(\ref{ljubav}) belongs to the set of quantum counterparts of
nonextensive entropies finding its application in increasing
number of problems in quantum physics, e.g. description and
controlling of laser cooling \cite{tannor}, a non-extensive
approach to the decoherence problem \cite{vidiella}, description
and quantifying of entanglement, and deducing criterions for
separability of density matrices \cite{horodecki,rajagopal}.

\section*{Conclusions}

In this work  we have stressed some conceptual difficulties
arising when Shannon's notion of information is applied to define
information gain in a quantum measurement. In particular we find
that the axiomatic derivation of Shannon's measure of information
is not applicable in quantum measurements in general. We also show
that the information content of a quantum system defined according
to Shannon's measure possesses some strongly non-physical
properties. We argue that these difficulties in defining the
information gain in quantum measurement by the Shannon measure of
information arise whenever it is not possible, not even {\it in
principle}, to assume that attributes observed are assigned to the
quantum system before the observation is performed.

Having critized Shannon's measure of information as being not
appropriate for identifying the information gain in quantum
measurement we proposed a new measure of information in quantum
mechanics that both mathematically and conceptually differs from
Shannon's measure of information. While Shannon's information is
applicable when measurement reveals a preexisting property, the
new measure of information takes into account that for quantum
systems the only features known before an experiment is performed
are the probabilities for various events to occur. In general,
which specific event occurs is objectively random.

The total information content of a quantum system defined
according to the new measure of information as the sum of the
individual measures of information for mutually complementary
observations is invariant under unitary transformations. This
implies that the total information content of the system is
invariant under transformation from one complete set of
complementary variables to another and is conserved in time if
there is no information exchange with an environment.

\section*{Acknowledgment}
In the previous version we did not make proper full reference to
the work of J. Uffink \cite{uffink}. We would like to thank J.
Uffink for pointing out this inadequancy as well as an error in
our previous Eq. (1). We also thank C. Simon for helpful comments
and discussions. This work has been supported by the Austrian
Science Foundation FWF, Project No. F1506 and the US National
Science Foundation NSF Grant No. PHY 97-22614.

\narrowtext


\vspace{-0.5cm}

\begin{thebibliography}{99}

\vspace{-1cm}

\bibitem{shannon} C. E. Shannon, Bell Syst. Tech. J. {\bf27}, 379
(1948). A copy can be found at
www.math.washington.edu/$^\sim$hillman/\\Entropy/infcode.html
\bibitem{kochen} S. Kochen and E. P. Specker, J. Math. and Mech. \bf 17\rm, 59 (1967).
\bibitem{peres} A. Peres, {\it Quantum Theory: Concepts and Methods},
(Kluwer Academic Publishers, 1995).
\bibitem{caslav} \v C. Brukner and A. Zeilinger, Phys. Rev. Lett.
{\bf 83}, 3354 (1999), quant-ph/0005084
\bibitem{feinstein} A. Feinstein, {\it Foundation of Information
Theory}\\ (McGraw-Hill, N.Y., 1958) p. 17.
\bibitem{uffink} J. Uffink, \it Measures of Uncertainty and the Uncertainty Principle\rm,
PhD thesis (Utrecht, 1990).
\bibitem{blabla} As suggested in \cite{uffink}
this should be contrasted to the cases where the notion of
information refers to knowledge about an unknown parameter in a
probability distribution (R. A. Fisher, Proc. Camb. Phil. Soc.,
{\bf 22}, 700 (1925), reprinted in R. A. Fisher, {\it
Contributions to Mathematical Statistics}, Wiley, N.Y., 1950), or
the information for discriminating between two probability
distributions (S. Kullback, {\it Information Theory and
Statistics}, Wiley, N.Y., 1959), or the information that one event
provides about another event (I. M. Gelfand and A. M. Yaglom in
{\it Arbeiten zur Informationstheorie II}, edited by H. Grell,
Deutscher Verlag der Wissenschaften, Berlin, 1957 p. 7. Russian
original in Uspekhi Mat. Nauk., {\bf 11}, 3, 1957).
\bibitem{jaynes} E. T. Jaynes, Phys. Rev. \bf106\rm, 620 (1957).
\bibitem{faddeev} D. K. Faddeev in {\it Arbeiten zur Informationstheorie
I}, edited by H. Grell (Deutscher Verlag der Wissenschaften,
Berlin, 1957) p. 88. Russian original in Uspekhi Mat. Nauk., {\bf
11}, 227 (1956).
\bibitem{jaynes0.5} E. T. Jaynes, {\it Probability Theory: The Logic Of Science}. To our knowledge this book
is only available on the web: http://bayes.wustl.edu/etj/prob.html
\bibitem{schroedinger}  E. Schr\"odinger, Naturwissenschaften {\bf 23}, 807
(1935). Translation published in Proc. Am. Phil. Soc. {\bf 124},
323 and in {\it Quantum Theory and Measurement} edited by J. A.
Wheeler and W. H. Zurek, (Princeton University Press, New Jersay,
1983). A copy can be found at: www.emr.hibu.no/lars/eng/cat
\bibitem{pauli} W. Pauli, Die allgemeinen Prinzipien der
Wellenmechanik in {\it Handbuch der Physik}, Band V, 1 (Hrsg. S.
Fl\"{u}gge, Springer-Verlag, 1990).
\bibitem{bell} J. S. Bell, Physics (Long Island City, N.Y.) 1 (1964) 195.
\bibitem{ghz} D. M. Greenberger, M. Horne, A. Shimony and A. Zeilinger, Am. J. Phys.
{\bf 58}, 1131 (1990).
\bibitem{scully} M. O. Scully, B. G. Englert and H. Walther, Nature {\bf 351}, 111 (1991).
\bibitem{horne} M. A. Horne, A. Shimony and A. Zeilinger, Phys.
Rev. Lett. {\bf 62}, 2209 (1989). T. J. Herzog, P. G. Kwiat, H.
Weinfurther and A. Zeilinger, Phys. Rev. Lett. {\bf 75}, 3034
(1995).
\bibitem{jaynes1} E. T. Jaynes, Information Theory in {\it Statistical Physics},
Brandeis Summer Institute (W.A. Benjamin inc, New York, 1962).
\bibitem{hausladen} P. Hausladen, R Jozsa, B. Schumacher, M.
Westmoreland, and W. K. Wootters, Phys. Rev. A {\bf 54}, 1869
(1996).
\bibitem{ivanovic} I. Ivanovic, J. Phys. A {\bf 14}, 3241 (1981).
\bibitem{wootters} W. K. Wootters and B. D. Fields, Ann. of Phys. {\bf 191}, 363
(1989).
\bibitem{hardy} G. Hardy, J. E. Littlewood, and G. P\'{o}lya {\it Inequalities},
(Cambridge University Press, Cambridge, 1952).
\bibitem{tsallis} C. Tsallis, J. Stat. Phys. {\bf 52}, 479 (1988).
\bibitem{ranyi} A. R\'{a}nya, {\it Wahrscheinlichkeitsrechnung mit einem
Anhang \"{u}ber Informationstheorie} (Deutscher Verlag der
Wissenschaft, 1962).
\bibitem{phd} \v C. Brukner, {\it Information in Individual Quantum Systems},
PhD Thesis (Vienna, 1999)
\bibitem{lee} J. Lee and M. S. Kim, Phys. Rev. Lett. {\bf 84},
4236 (2000); J. Lee, M. S. Kim, Y. J. Park, and S. Lee,
quant-ph/0003060
\bibitem{tannor} A. Bartana, R. Kosloff and D. J. Tannor, J. Chem. Phys. {\bf 106}, 14358 (1997); D.
J. Tannor and A. Bartana, J. Phys. Chem. {\bf 103}, 10359 (1999).
\bibitem{vidiella} A. Vidiella-Barrenco and H. Moya-Cessa,
quant-ph/0002071.
\bibitem{horodecki} R. Horodecki, M. Horodecki,
Phys. Rev. A {\bf 54}, 1838 (1996). R. Horodecki , P. Horodecki,
and M. Horodecki, Phys. Rev. Lett. A {\bf 210}, 377 (1996).
\bibitem{rajagopal} S. Abe and A. K. Rajagopal, Physica A {\bf 289}, 157 (2001);
Chaos, Solitons, and Fractals, {\bf 13}, 431 (2002), C. Tsallis,
S. Lloyd, and M. Baranger, e-print: quant-ph/0007112.
\end{thebibliography}
\end{document}